\documentclass{eas}
\usepackage{graphicx}
\begin{document}

\title{Monitoring evolved stars for binarity\\
with the HERMES spectrograph} 
\runningtitle{HERMES binary survey}
\author{N. Gorlova}\address{Institute of Astronomy, KUL, Belgium}
\author{H. Van Winckel}\sameaddress{1}
\author{J. Vos}\sameaddress{1}
\author{R. H. {\O}stensen}\sameaddress{1}
\author{A. Jorissen}\address{Institute of Astronomy and Astrophysuics, ULB, Belgium}
\author{S. Van Eck}\sameaddress{2}
\author{N. Ikonnikova}\address{Sternberg Astronomical Institute, Russia}
\begin{abstract}

Binarity is often invoked to explain peculiarities
that can not be explained by the standard theory of stellar evolution.
Detecting orbital motion via the Doppler effect
is the best method to test binarity when direct imaging is not possible.
However, when the orbital period exceeds length of a typical observing run,
monitoring often becomes problematic.
Placing a high-throughput spectrograph on a small semi-robotic telescope
allowed us to carry out a radial velocity survey
of various types of peculiar evolved stars.
In this review we highlight some findings after the first four years of observations.
Thus, we detect eccentric binaries among hot subdwarfs, barium, S stars,
and post-AGB stars with disks, which are not predicted by the
standard binary interaction theory.
In disk objects, in addition, we find signs of the on-going mass transfer
to the companion, and an intriguing line splitting, which we
attribute to the scattered light of the primary.

\end{abstract}
\maketitle
\section{Unique binary survey enabled by HERMES on Mercator}

HERMES is the state-of-the-art fiber echelle spectrograph attached to the Flemish 1.2-m telescope
Mercator on La Palma
(Raskin {\em et al.\/} \cite{Raskin2011}).
Our survey is based on the use of a high-resolution mode ($R=85,000$).
High throughput allows to observe stars up to $V=14$ (S/N$\sim$20 in 1 hr),
while temperature-controlled environment allows radial velocity (RV) determination
with precision better than 200 m s$^{-1}$. Spectra are reduced with the \textsc{python}-based pipe-line,
and RVs are determined by cross correlation using line masks adapted
for a range of spectral types and metallicities. 
The survey of post-Main sequence (MS) stars is the largest program
that has been continuously run since the commissioning of HERMES in 2009.
The observations are carried out in the queue mode,
by the HERMES consortium members led by the Catholic University of Leuven.
Stars are observed with frequencies ranging from once per week to several times per semester,
depending on the expected time-scale of variability.
Such mode of operation allows to collect a unique time series of spectra.

The main goal of the survey is to verify, by means of the RV monitoring,
binarity (or the absence of it in the control samples)
among all main groups of post-MS stars where it was invoked
to explain various peculiarities: post-asymptotic giant branch (post-AGB) stars with hot and cold dust, central stars of planetary nebulae (PNe),
silicate J-type stars, subdwarfs, symbiotics, chemically peculiar giants, R CrB and W Ser types, and several others
(see Van Winckel {\em et al.\/} \cite{VanWinckel2010}).
Some of these classes must reflect different evolutionary stages
of the same systems, but the empirical evidence for it is largely missing.
By characterizing orbits, stellar properties, and dynamics of the circumstellar matter,
we hope to fill in this gap.
In this contribution we present some major findings of the first four years of the survey
and their implications for the theory of binary evolution.

\section{Highlight of some discoveries}

\subsection{Long-period subdwarfs}

We have been monitoring 16 hot subdwarfs (types sdB and sdO) with HERMES.
Targets were selected based on the presence of lines from a cool (F--K) companion in their spectra.
Hot subdwarfs are core helium burning objects that have lost a considerable fraction of their hydrogen atmospheres.
Theory predicts two main channels producing these objects, both involve binary systems
(Han {\em et al.\/} \cite{Han2002}, \cite{Han2003}).
The first channel is via a common envelope ejection, which should result in the short-period orbits (P$<$10 d).
Such systems have been already extensively described in the literature. The second channel is via a stable Roche-lobe overflow (RLOF),
which occurs for more massive companions. In the latter case predicted orbital periods
can be as long as 500 days, but no such systems have been known when we started our monitoring. 

Until now we discovered binarity in 8 systems with periods between 491 and 1363 days
({\O}stensen \& Van Winckel \cite{Oestensen2011}, \cite{Oestensen2012}; Vos {\em et al.\/} \cite{2012Vos},
\cite{2013Vosa}, \cite{2013Vosb}).
Radial velocities for the cool companions were obtained from the cross-correlation function (CCF) with the standard HERMES masks,
while for the subdwarfs themselves a cross-correlation with a synthetic spectrum around He\,\textsc{i} line at 5876 \AA\,
was performed. By fitting an orbital solution independently to both components' RV curves,
we discovered for the first time a difference of $\sim$1 km/s between the two systemic velocities,
as can be expected due to the larger gravitation redshift at the surface of a compact subdwarf.
Surprisingly, for the majority of systems the periods turned out to be longer and eccentricities higher
than predicted by the RLOF theory.
Similar effect is observed also in AGB and post-AGB systems, that have undergone mass exchange
further along the evolutionary sequence (see the next Sections).

\subsection{The eccentricity -- period diagram of barium and S stars}

Barium stars are a class of K giants with strong lines of elements like barium 
produced by the s-process of nucleosynthesis. 
These overabundances result from the mass transfer in a binary system, the polluting 
heavy elements being formerly produced within an AGB companion, now a very faint white dwarf. The exact mode of mass transfer responsible for that pollution remained uncertain though, mainly because many barium systems are found in rather narrow orbits which could not accomodate an AGB star (Pols {\em et al.\/} \cite{Pols2003},
Izzard {\em et al.\/} \cite{Izzard2010}, and references therein). To fully constrain the mass-transfer mechanism, it is important to obtain an eccentricity -- period diagram for a {\it complete} sample of barium stars. In this way, it is possible to evaluate the limits in orbital periods defining the barium syndrome (the lower limit is important to distinguish between  
stable and unstable RLOF, whereas the upper limit is important to infer the efficiency of mass accretion through stellar winds).
There is a number of systems where long periods were not covered yet.
With the advent of the HERMES spectrograph, that goal may finally be reached, since the time coverage now amounts to more than 30 years when combining the older CORAVEL or DAO measurements with the new HERMES ones (the CORAVEL and HERMES measurements are both in the IAU system of radial velocities; Udry {\em et al.\/} \cite{Udry1999}).  

Here we report on the results of the CORAVEL/HERMES monitoring of the complete sample of 34 barium stars with strong anomalies (classified as Ba3 to Ba5 on Warner's 1-5 scale; Warner \cite{Warner1965}) from the list
of Lu {\em et al.\/} (\cite{Lu1983}), and a sample of 33 barium stars with mild anomalies (Ba1 - Ba2), randomly selected. The RV monitoring was extended to a sample of bright,
northern S stars from the General Catalogue of Galactic S Stars
(GCGSS; Stephenson \cite{Stephenson1984}) with no variable star designation, since Jorissen {\em et al.\/} (\cite{Jorissen1993}, \cite{Jorissen1998}) suggested that S stars without lines from the unstable element Tc are the cooler analogs of barium stars. Those three samples are the same as those monitored by Jorissen {\em et al.\/} (\cite{Jorissen1998}).  At this point, we can conclude that {\it all\footnote{With the exception of the southern target HD~19014 for which no further HERMES data could be secured, so that the question of its binarity remains pending.} the monitored targets turn out to be binaries}, and (preliminary) orbits are available for all of them (but HD~50843, HD~65854, HD~95345, HD~104979 and HD~184185, all of which nevertheless showing clear radial-velocity variations; in the 1998 study, they were flagged as having no or weak signature of duplicity).   
The precision of the HERMES spectrograph is such that it allowed us to derive orbits with mass functions as small as $2\times10^{-4}$~M$_{\odot}$ (HD~18182).

\begin{figure}[ht!]
 \begin{center}
 \includegraphics[width=7cm]{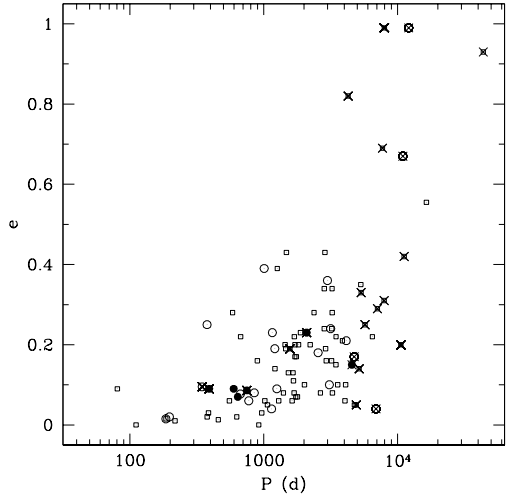}
 \end{center}
 \caption[]{\label{Fig:elogP}
The eccentricity -- period diagram of barium stars (squares) and Tc-poor S stars (circles). The crosses identify those orbits obtained thanks to HERMES data. The black circles denote symbiotic S stars identified
from their H$\alpha$ emission (see Van Eck \& Jorissen \cite{VanEck2002}).   
}
\end{figure}

The resulting eccentricity -- period ($e - P$) diagram is shown on Fig.~\ref{Fig:elogP}.
There are several noteworthy features:
\begin{itemize}
\item The HERMES data obviously contributed to filling the $e - P$ diagram at the longest periods and largest eccentricities. Among these long-period orbits, only HD~123949 ($P = 7975\pm73$~d, $e=0.99\pm0.10$) and HD~196673 ($P = 7721\pm131$~d, $e=0.69\pm0.04$) are well constrained, because the periastron passage is covered. Nevertheless, the $e - P$ diagram indicates for the first time without doubts that {\it barium stars are not found in systems with periods in excess of about $2\times10^4$~d}. This is an important result to constrain the efficiency of accretion through wind mass transfer.       
\item Tc-poor S stars and Ba stars behave exactly the same in the $e - P$ diagram (except may be for the absence of circular S systems in the period range 300 -- 1000~d, and their dominance among Ba systems), confirming once more that their difference is just one of surface temperature, but not of binary-evolution channel.
\item the gap at $e < 0.05$ and $P > 10^3$~d is well apparent\footnote{The orbital elements for the system HD~49641, as computed by McClure \& Woodsworth (\cite{McClure1990}), make it fall in the gap; however, the errors on the velocities are such that a value $e = 0.07$, as used here, is possible as well.} and is reminiscent of the situation prevailing among pre-MS and MS systems (Mathieu \cite{Mathieu1992}, \cite{Mathieu1994}). It is somewhat narrower among our sample of post-mass-transfer systems though.  
\item there appears to be a well-defined upper limit on the eccentricity at a given orbital period, and the systems defining this threshold are somewhat separated from the bulk of the sample, restricted to $e < 0.3$. One may wonder whether or not these two groups correspond to different evolutionary channels.    
\end{itemize}
The $e - P$ diagram of barium and Tc-poor S stars as it stands now, in its almost final form, will undoubtedly constitute a great challenge for binary evolution models.

\subsection{Post-AGB stars with disks}

There are $\sim$70 high-altitude supergiants in our sample. Because of their isolated location
and dust shells they are thought to be in the post-AGB stage.
In about half of them the shells are expanding and must be remnants of the AGB winds.
In the other half, however, dust resides in disks,
which is more difficult to explain. The leading hypothesis today
is that disk objects are binary systems that have exchanged mass in the past.
Once the current primary becomes a white dwarf, and the presumably MS companion
a red giant, the pair may emerge as a Barium/S-type system,
provided that the primary evolved all the way through the AGB phase
and transferred the synthesized material onto the companion.
This effect, however, may be smeared due to the accretion of the re-processed gas from the disk.
Ten more objects have no or very little dust.
They were included in the survey because of the RV Tau type of variability
or depletion in refractory elements -- characteristics shared with other post-AGB stars. 
By monitoring RVs, we want to test whether disk objects, as well as depleted ones are all binaries,
while shell and dust-free, normal-composition objects are mostly single (Van Winckel \cite{VanWinckel2003}).
Here we report first results for the disk and the dust-free sub-samples. 

The most recent previous study on this subject was done by Van Winckel {\em et al.\/} (\cite{VanWinckel2009}),
who demonstrated binarity for all 6 disk systems in their sample.
Stars in that study were relatively weak pulsators.
Pulsations introduce scatter in the orbital RV curve --
the larger the amplitude and the period of pulsations,
the more difficult it is to study the orbital motion.
The goal of our survey is to characterize all known binary post-AGB candidates up to the 12$^{th}$ mag
in the Northern hemisphere,
independent of the pulsation properties.

\begin{figure}[h!]
 \includegraphics[width=7cm]{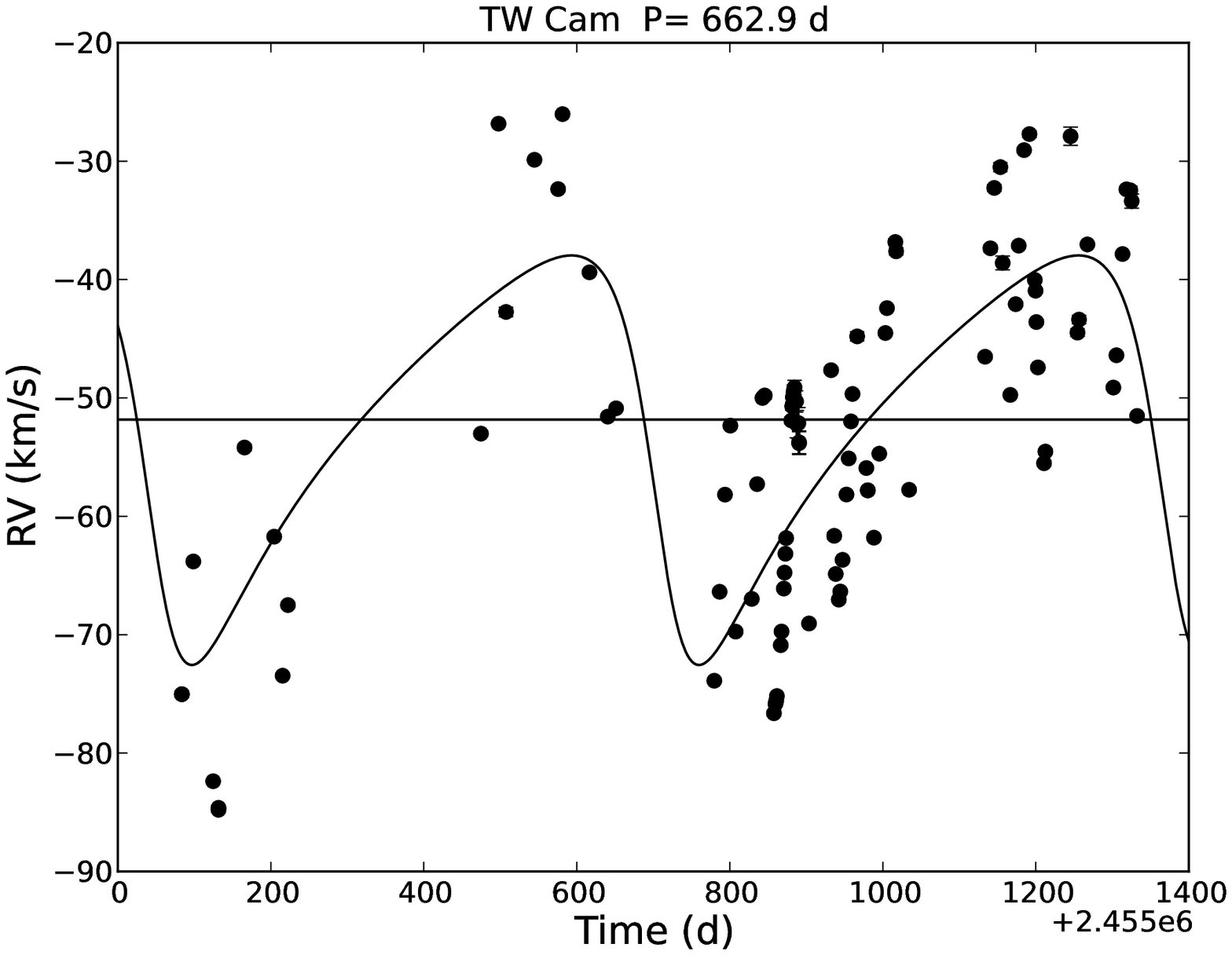}
 \hspace*{-1.5cm} 
 \qquad
 \includegraphics[width=7cm]{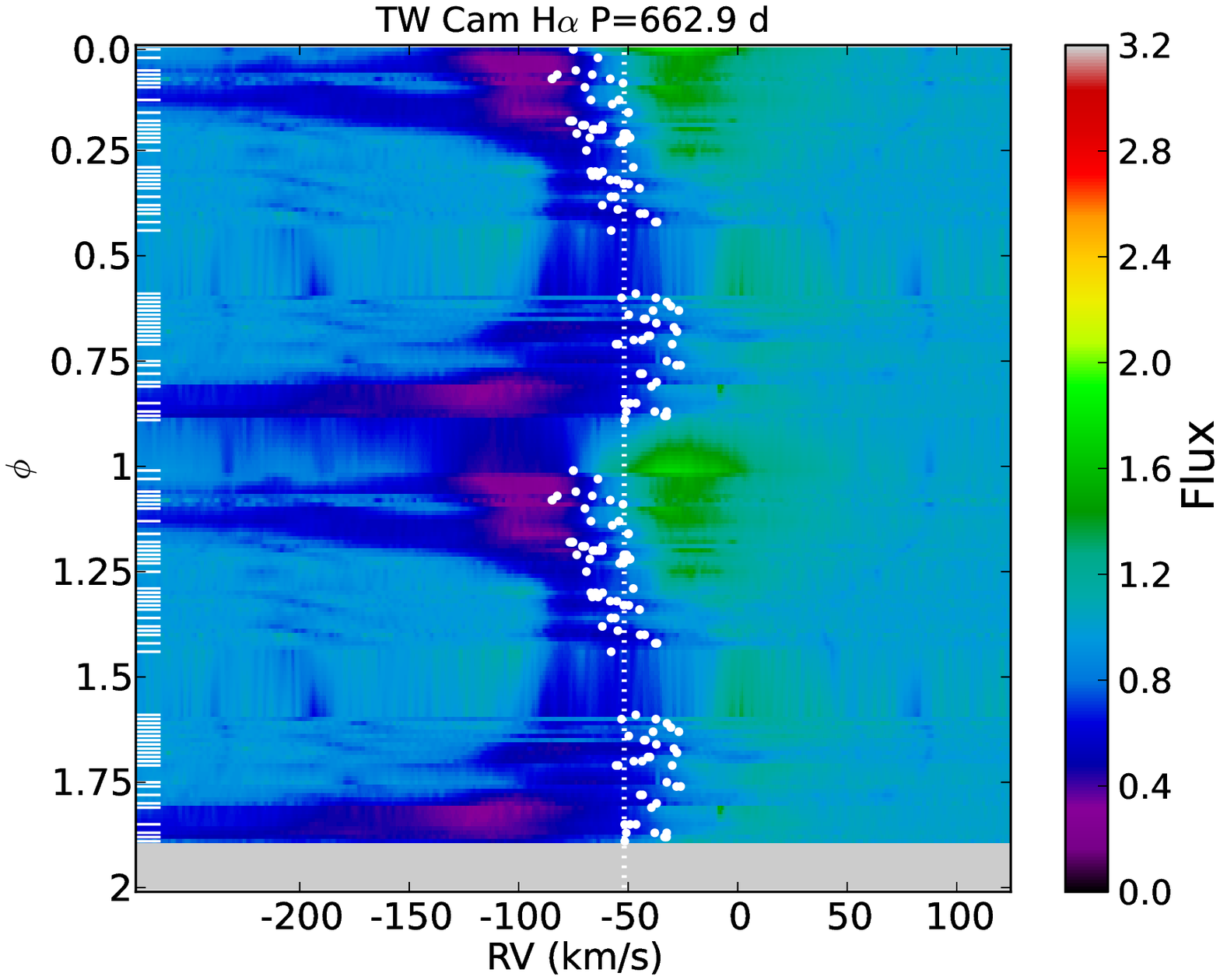}
 \caption[]{\label{Fig:TWCamHal}
TW Cam -- one of the RV Tau-type pulsating stars with disks that we discovered to be binary.
{\it Left panel:} RV curve fitted with a Keplerian orbit (P $=$ 662.9 d, $e=0.44$).
The 'scatter' is due to pulsations with P$\sim$80 d.
{\it Right panel:} trailed spectrogram of H$\alpha$ plotted against the orbital phase
(time runs down, one period is shown twice to guide the eye,
vertical line designates systemic velocity, white dots -- photospheric velocities).
}
\end{figure}

So far we have detected long-term RV variations
in 22 out 27 disk objects, and only in 3 out of 13 weak/no-disk ones,
consistent with the disk-binarity paradigm. The best binary candidates with complete periods discovered in our survey are:
IRAS 19135$+$3937 (P$=$127 d), BD$+$46 442 (P$=$141 d) (Gorlova {\em et al.\/} \cite{Gorlova2012a}, \cite{Gorlova2012b});
IRAS 11472$-$0800 (P$=$638 d)  (Van Winckel {\em et al.\/} \cite{VanWinckel2012}),
TW Cam (P$=$663 d, Fig. \ref{Fig:TWCamHal}), DF Cyg (P$\sim$775 d),
V Vul  (P$=$700--800 d), CT Ori (P$\sim$1000 d), EP Lyr (P$\sim$1100 d),
R Sge (P$=$1159 d),  and RV Tau (P$\sim$1210 d). 
In some of these objects similar photometric periods have been reported,
the reason for the RV Tau {\it b} or the semi-regular (SR) classification.
These slow brightness variations were ascribed to either beating modes of pulsations
or to dust production events similar to those in R CrB stars.
Binarity provides a more simple explanation of this phenomenon:
the primary is seen at different elevations above the disk plane
as it travels along the orbit, resulting in the variable extinction.

In many disk systems we discovered a specific behavior in H$\alpha$
that does not correlate with pulsations, but rather with the long-term RV variations.
In these systems H$\alpha$ shows a double-peak emission profile, that
becomes asymmetric or is entirely replaced, by a wide blue-shifted absorption
during primary's superior conjunction.
Few more systems show a permanent P Cyg-like profile that may intensify at the same phase. 
This effect was first noticed in non-variable objects,
and here we report the detection of it in the pulsating objects as well (see example in Fig. \ref{Fig:TWCamHal}).
Following Thomas {\em et al.\/} (\cite{Thomas2013}) who modeled this behavior in the Red Rectangle nebula, 
we attribute it to the on-going mass transfer from the giant primary to the companion.
The double-peak emission likely originates in the accretion disk around the companion,
while the blue absorption -- in one of the jet's lobes emanating from the disk, when it
crosses our line of sight to the giant (Gorlova {\em et al.\/} \cite{Gorlova2012a}).
Systems with permanent P Cyg profiles then must be seen closer to pole-on.
The fact that the accretion is still going on at this evolutionary stage may have implications for the
evolution of the orbit, formation of the chemical anomalies on either of the components,
and the appearance in the PN stage.

\begin{figure}[hb!]
 \begin{center}
 \includegraphics[height=8cm]{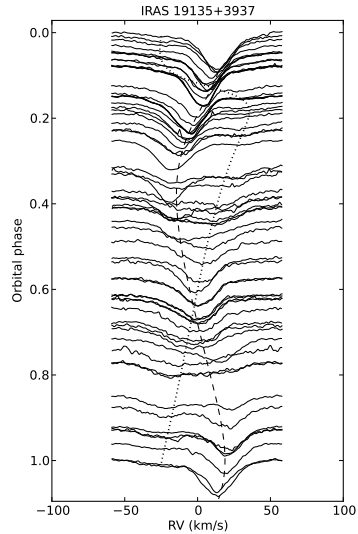}
 \end{center}
 \caption[]{\label{Fig:iras191ccf}
Cross-correlation function of IRAS 19135$+$3937, a post-AGB star with a disk,
plotted against the orbital phase.
Dashed line designates velocity of the primary component, that we attribute to the
direct light from the post-AGB star,
while dotted line designates a satellite component,
that we attribute to the reflected light of the primary, rather than to a companion.
}
\end{figure}

Except for the gravitation pull on the primary, the companions remain elusive.
In Gorlova {\em et al.\/} (\cite{Gorlova2012a}) we detected a weak satellite feature
in the CCF of BD$+$46 442, and now we report it
for another similar object IRAS 19135$+$3937 (Fig. \ref{Fig:iras191ccf}). In RV Tau stars photospheric metal lines split
twice during pulsation period due to the propagation of shock waves. These two objects, however,
do not pulsate. Can it be a spectrum of the companion?
The similarity of spectra in the opposite conjunctions
indicates very similar spectral and luminosity types between the primary and the companion.
Having a post-MS companion to a post-AGB star is, however, highly unlikely,
due to the short duration of this evolutionary phase. The secondary component
might be instead light of the primary reflected off the inner wall of the circumbinary disk.
This explanation is supported by our photometry of IRAS 19135$+$3937,
that shows that the object becomes bluer during minimum light (Gorlova {\em et al.\/} 2014, in prep.),
as well as by the recent interferometric study of a prototype object 89 Her, where Hillen {\em et al.\/} (\cite{Hillen2013})
found up to 40\% of the optical flux to result from scattering.

\section{Summary and future work}

Within the HERMES survey of evolved stars we detected
many suspected binaries,
characterized their orbits, abundances, and the circumstellar environment.
In this contribution we presented evidence for the existence of
eccentric and long-period systems that can not be explained by the standard binary interaction theory.
In post-AGB stars with disks we detected in addition
an active mass transfer and a scattered light component in the spectra.
We plan to employ Doppler tomography and interferometry to follow
up on these spectroscopic discoveries.


\end{document}